\documentclass[%
 reprint,
 amsmath,amssymb,
 aps,
]{revtex4-2}
\usepackage{amsmath,amssymb,amsfonts}
\usepackage{slashed}
\usepackage{geometry}
\usepackage{graphicx}
\def\sech{\mathrm{sech}}

\def\sech{\mathrm{sech}}

\newcommand{\sfrac}[2]{{\textstyle\frac{#1}{#2}}}

\usepackage[colorlinks]{hyperref}
\hypersetup{
    colorlinks=true,
    linkcolor=blue,
    filecolor=magenta,  
    urlcolor=[rgb]{0.00,0.00,0.50},    
    citecolor=magenta,
    }
\usepackage{url}

\begin{document}
%
%\preprint{APS/123-QED}

\title{Integrability meets the charged relativistic particle}

\author{Francisco Correa$^a$}
\email[Electronic mail: ]{francisco.correa@uach.cl}
\author{Justo L\'opez-Sarri\'on$^b$}
\email[Electronic mail: ]{justo.lopezsarrion@ub.edu}
\affiliation{$^b$Instituto de Ciencias F\'isicas y Matem\'aticas, Universidad Austral de Chile, Casilla 567, Valdivia, Chile}
\affiliation{$^b$Departament de F\'\i sica Qu\`antica i Astrof\'\i sica and 
Institut de Ci\`encies del Cosmos (ICCUB), \\ Universitat de Barcelona, 
Mart\'i  Franqu\`es 1, 08028 Barcelona, Spain}

%\affiliation{
%\affiliation{
%\affiliation{%


%             %  but any date may be explicitly specified

\begin{abstract}
We notice an analogy between the motion of a relativistic particle with external homogeneous and time-dependent electromagnetic fields and the Dik'ii-Eilenberger equation for the Bogoliubov-de Gennes equation. By means of the integrable defocusing nonlinear Schr\"odinger hierarchy and their solitons appearing in the Gross-Neveu and Nambu-Jona-Lasinio models, we construct an infinite family of solutions of the Lorentz equation with non-vanishing curvature and torsion.
\end{abstract}

\maketitle

The study of the electromagnetic interaction of a relativistic point particle of mass $m$ and charge $e$ appears as a standard textbook example of classical and quantum field theories \cite{books1, books2} and finds current applications in several research areas, for example, in high-power laser physics with strong electromagnetic fields \cite{laser} and plasma physics \cite{plas1,plas2}.  Neglecting radiative terms, the classical Lorentz equation in terms of the four-velocity vector $\eta^\mu$ as a function of the proper time $\tau$ reads
  \begin{equation} \label{lorentzp}
m \frac{d\eta^\mu}{d\tau}=\frac{e}{c} F^\mu_\nu \eta^\nu \, .
\end{equation}
The solutions of the relativistic Lorentz equation have been studied using various approaches. One of them is the Frenet-Serret formalism, which is coordinate independent and carries the direct geometric description of the word line \cite{books2, fsf}.  Other methods used in the literature, such as those inspired by the computation of Dirac propagators in the presence of certain external fields \cite{gm}, work mainly for homogeneous fields, i.e. with a covariantly constant field tensor \cite{synge}. More recently, the well-known solvability of a relativistic particle in a plane wave electromagnetic background has been understood in terms of integrability \cite{hiint} and extended to scalar field backgrounds \cite{ahi}. The advantages of integrable models have been widely tested in interacting fermion theories \cite{integrableGN, shei}, gravity \cite{gravsol, gravityprl} or Yang-Mills self-dual theories \cite{YM}, just to name a few. As a common ingredient among the physical applications of integrable equations, the nonlinear Schr\"odinger (NLS) hierarchy has played a significant role in optics, superconductivity, plasma physics \cite{nlsbook}, non-Hermitian physics \cite{nonlocal} and it is also related with the Frenet-Serret and vortex solutions \cite{hasimoto}.   From the Lax pair or zero curvature formalism, the associated linear problems of the NLS hierarchy involves the Bogoliubov-de Gennes (BdG) Hamiltonian \cite{bdgh}. This relationship has been used to solve the gap equation and to find non-trivial solutions of the Gross-Neveu and Nambu-Jona-Lasinio models \cite{bdg1, sols}, clarifying the integrable features of the Ginzburg-Landau expansion \cite{bdg2}. In this article, we investigate an analogy between the stationary nonlinear Schr\"odinger hierarchy and the motion of a classical charged relativistic particle under homogeneous but time dependent electromagnetic fields. The key element for this connection is the diagonal resolvent expansion of the Bogoliubov-de Gennes (BdG) Hamiltonian that satisfies the Dik'ii-Eilenberger equation, known in the context of superconductivity \cite{eilen, super} and fermionic theories \cite{bdg1, bdg3}. We show how this equation can be mapped exactly in the dynamical equations (\ref{lorentzp}) of a relativistic particle allowing to translate the infinite class of solutions of the integrable hierarchy into trajectories under certain electromagnetic fields. The layout of the article is the following. In Section \ref{sec1}, the Lorentz equation of a relativistic particle is recast in a suitable matrix form. In Section \ref{sec2}, the correspondence between the Dik'ii-Eilenberger and the Lorentz equations is established together with their solutions in terms of the diagonal resolvent. Several examples of relativistic trajectories and their geometric properties are illustrated in Section \ref{sec3}. The Section \ref{sec4} is devoted for the discussion and outlook.  \vspace{-1em}

 \section{A charged relativistic particle}\label{sec1}

The setup consists of a charged particle moving in uniform electromagnetic fields, where the magnetic field is constant pointing the $z-$axis and a perpendicular time-dependent electric field. Due to the axial symmetry, we can focus mainly on the planar motion in (2+1)- dimensions, bearing in mind that the extension to the $z$-direction is straightforward. The dimensional reduction leads to the same  Lorentz equation (\ref{lorentzp}), but with a Minkowski metric in $g^{\mu\nu}=\mathrm{diag}(+,-,-)$ and the  indices $\mu$ y $\nu$ running from  0 to 2. The covariant velocity has a different normalization as a remainder of the reduced dimension,
\begin{equation}\label{norm}\eta^\mu\eta_\mu=(\eta^0)^2 -(\eta^1)^2-(\eta^2)^2=1+(\eta^3)^2 \, , 
\end{equation}
where $\eta^3$ is the constant component of the four velocity in the $z-$direction, so without loss of generality the r.h.s of (\ref{norm}) can be fixed to the unity. 
By setting the  natural units $e=m=c=1$ we write (\ref{lorentzp}) as $\eta^{\prime\mu}= F^{\mu\nu}\eta_\nu$, with the prime meaning derivative with respect to $\tau$. Next, we consider the skew-symmetric  tensor $F^{\mu\nu}$, 
 \begin{equation}\label{ftensor}
 F^{\mu\nu} \equiv \left(\begin{matrix}
0&-E_1&-E_2\\
E_1&0& B\\
E_2&-B &0\end{matrix}\right)\, ,
\end{equation} 
in terms of  the constant magnetic field $B$ transversal to the plane and the components of the time varying, but uniform, electric fields $E_1=E_1(x^0)$ and $E_2=E_2(x^0)$. 
Although the fields are functions of $x^0$, they can be evaluated on the particle trajectory such that depend only on the particle proper time $\tau$ through the relation $x^0=x^0(\tau)$. Thus, given a solution of the motion equation (\ref{lorentzp}), it is equivalent to write the fields as functions of $x^0$ or $\tau$.  
As we shall see below, it will be convenient from now on, to write the electric field components as functions of $\tau$,
$E_1(\tau)=E_1(x^0(\tau))$ and $E_2(\tau)=E_2(x^0(\tau))$ \cite{porsi} and group them in a complex function as,
\begin{equation}
\mathcal{E}=E_1+iE_2\,.
\label{complexElectric}
\end{equation}  
In this way, it is possible to write (\ref{ftensor}) in terms of its dual  vector $G^\mu$ defined via $F^{\mu\nu} \equiv 2\epsilon^{\mu\nu\rho} G_\rho $ .   The vectors 
  \begin{equation}
  \eta^\mu=(\sfrac{dx^0}{d\tau},\sfrac{dx^1}{d\tau},\sfrac{dx^2}{d\tau}), \quad 
  G^\mu =\sfrac12 (B,E_1,E_2)\, ,
  \end{equation}
can be promoted into matrix form by means of the  $\gamma$ matrices, $\gamma^0=\sigma^3,\gamma^1=i\sigma^2,\gamma^2=i\sigma^1$ which satisfy the  Clifford algebra, $\left\{\gamma^\mu,\gamma^\nu\right\}=2g^{\mu\nu}$ and also  $  \left[\gamma^\mu,\gamma^\nu\right]=2i\epsilon^{\mu\nu\rho}\,\gamma_\rho$. Here we denote $\sigma$ as the standard Pauli matrices and $\epsilon^{012}=+1$. By virtue of this algebra we obtain,
$$i\left[\slashed G,\slashed\eta\right]=-2\epsilon^{\mu\nu\rho} G_\mu\eta_\nu\gamma_\rho= (F^{\mu\nu}\eta_\nu)\gamma_\mu \, ,$$ which, in terms of  $\gamma^\mu\eta_\mu=\slashed{\eta}$ and $\gamma^\mu G_\mu =\slashed{G}$ allows to write down the Lorentz equation in the following way,
\begin{equation}
\slashed{\eta}^\prime = i\left[\slashed G,\slashed\eta\right]=\left[\left(\begin{matrix}
\sfrac{1}{2} B&-\sfrac{1}{2}\mathcal{E}(\tau)\\[1.5 pt]
\sfrac{1}{2}\mathcal{E}^*(\tau)&-\sfrac{1}{2}B\end{matrix}\right),\slashed{\eta}\right]\ \, .
\label{eilen1}
\end{equation}
This matrix representation of the Lorentz equation is the first step to link the current setup to a whole hierarchy of nonlinear integrable equations.  \vspace{-1em}
 
 \section{Lorentz and Dik'ii-Eilenberger}\label{sec2}

The Lorentz equations written as the Eq.  (\ref{eilen1}) have the exact same form of the Dik'ii-Eilenberger equation for the resolvent operator $R$,  
 \begin{equation}
R(\tau;B/2)\equiv \langle\tau\vert  (H-(B/2))^{-1}\vert \tau\rangle=\slashed{\eta}\sigma^3\, ,
 \label{resol}
 \end{equation}
associated with the following BdG Hamiltonian where the constant magnetic field $B$ plays the role of the energy eigenvalue,
\begin{equation}
H=\left(\begin{matrix}
-i\sfrac{d}{d\tau}&\frac{1}{2}\mathcal{E}(\tau)\\[2pt]
 \sfrac{1}{2} \mathcal{E}^*(\tau)& i\sfrac{d}{d\tau}
\end{matrix}\right)\quad , \quad H\psi=\frac{B}{2}\psi \quad .
\label{hamiltonian}
\end{equation}

In this way, the matrix resolvent of the BdG Hamiltonian contains in its entries the components of a relativistic charged particle. Although it is surprising how these equations can be related to each other, finding an analytical solution to the Dik'ii-Eilenberger equation in the general case, as well as to the Lorentz equation, is not an easy task.  However, for certain choices of electric fields, the energy expansion of the resolvent can be found analytically. In the integrable language the Lorentz equation in the form (\ref{eilen1}) is nothing else than the stationary version of the zero curvature formalism of the defocusing hierarchy of nonlinear Schr\"odinger equations denoted by  nS$^{m}$ \cite{solitons}. In other words,  the problem for the trajectories can be solved completely if the electric fields as (\ref{complexElectric}) are a solution of any of the nS$^{m}$ equations. Explicitly, this can be seen considering the resolvent expansion in terms of the electromagnetic fields
\begin{eqnarray}\label{resolvent}
    R(\tau)=\begin{pmatrix} \eta^0 & \eta^+\\ \eta^- & \eta^0 \end{pmatrix}=\sum_{\ell=0}^{m+1} \left(\sfrac{B}{2}\right)^{m+1-\ell }  \begin{pmatrix}
 g_\ell & f_{\ell-1} \\[2pt]
   f_{\ell-1}^* & g_\ell 
    \end{pmatrix} \, ,
\end{eqnarray}
where $\eta^{\pm}=\eta^1\pm i\eta^2$ and $g_\ell(\tau)$ and $f_\ell(\tau)$ are functions defined recursively by \cite{solitons}
\begin{align}\label{recf}
  f_\ell&=-\frac{i}{2}f_{\ell-1}^{\,\prime}{+}(\mathcal{E}/2)  g_\ell , \\ \label{recg}
  g_\ell&=\int i\left((\mathcal{E}^*/2)f_{\ell -1}{-}(\mathcal{E}/2)\, f_{\ell -1}^*
    \right){+} c_\ell  \, .
\end{align}
with $f_{-1}=0, f_0=\mathcal{E}/2, g_0=1,g_1=c_1$ and the constants $c_\ell$ are given in terms of the spectral data of the associated BdG equations \cite{constants}. For each $m$, the recursion expansion (\ref{resolvent}) is truncated if ${\cal E}$ satisfies the $\text{nS}^m$ equations which are given as $f_{m}=f_{m}^*=0$ from (\ref{recf}). The first members of the hierarchy with $m=0,1,2$
\begin{align}\label{chiral}
\text{nS}^0:& \, \mathcal{E}'{+}2ic_1\mathcal{E}=0 &\\ \label{nls}
\text{nS}^1:& \,  \mathcal{E}''{-}2|\mathcal{E}|^2\mathcal{E}{+}2i c_1\mathcal{E}'{-}4c_2\mathcal{E}=0 &\\ \notag
\text{nS}^2:& \,  \mathcal{E}'''{-}6|\mathcal{E}|^2\mathcal{E}'{+}2ic_1(\mathcal{E}''{-}2|\mathcal{E}|^2\mathcal{E}){-}4c_2\mathcal{E}'{-}8ic_3\mathcal{E}=0& 
\end{align}
are known as the chiral, nonlinear Schr\"odinger and Hirota equations respectively. When $\mathcal{E}$ is real the stationary Hirota  equation is reduced to the modified Korteweg-de Vries equation \cite{solitons}. Once this analogy is established, several physical consequences can be paired from both sides. First, from the resolvent expansion in the matrix form (\ref{resolvent}), it is possible to check that the hierarchy nS$^{m}$, and therefore the Lorentz equation, displays an exact $U(1)$ symmetry on the electric fields combinations (\ref{complexElectric}), which are defined up to a constant phase $\widetilde{\mathcal{E}}=e^{i \phi}\mathcal{E}$. This means that we can design new electric field configurations where the trajectories are rotated by an  angle $\phi$ in the plane perpendicular to $B$. Second, as a well-known feature of integrable hierarchy of equations, once the electric fields combination $\mathcal{E}$ satisfy one of the equations of the hierarchy $\text{nS}^m$, the infinite tower higher order equations are automatically solved. This is done by finding the coefficients $c_{j}, j>m$ or the infinite number of constants of motion in terms of the trajectories and electromagnetic fields given by the condition $g_j=0, j>m$ from the recursion relation (\ref{recg}). Since the nonlinear equations are functions of the proper time and we are dealing with the stationary hierarchy, there is no an additional dynamical parameter and it becomes natural do not expect the standard interpretation of such conversed charges, in contrast with \cite{hiint, ahi}. Instead, they appear in the curvature ${\cal K}$ and torsion ${\cal T}$ of those world-lines generated by the electromagnetic background and are helpful to simplify their form, as we shall see in the examples of the next section. The Frenet-Serret equations take the form
\begin{equation}
\dot{\eta}^\mu={\cal K}\,k^\mu, \quad \dot{k}^\mu={\cal K}\,k^\mu -{\cal T} b^\mu, \quad \dot{b}^\mu={\cal T} k^\mu \, ,
\end{equation}
where we defined the normal,  $k^\mu$, and binormal vector $b^\mu$. The curvature takes the form ${\cal K}^2=\eta_\mu F^{\mu\nu} F_{\nu\alpha} \eta^\alpha=(\eta^+\mathcal{E}^*{+}\eta^-\mathcal{E}{-}B \eta^0)^2{-}B^2{-}|\mathcal{E}|^2 $ while the torsion ${\cal T}=b\cdot\dot{k}=b^\mu\dot{k}_\mu$.  The normalization condition $\eta^\mu\eta_\mu=1$ is provided by integrability and the spectral polynomial of the BdG quantum problem (\ref{hamiltonian}). Using the hyper-elliptic curve ${\cal N}_m(B)=\prod_{i=0}^{2m+1}(B-B_i)$ of the associated defocusing nonlinear Schr\"odinger hierarchy one can always choose a multiplicative constant in (\ref{resolvent}) such that $\det R=1$. Each solution $\mathcal{E}$ of the integrable equation defines simultaneously a word-line and a quantum BdG problem with eigenvalue $B/2$ such that the constants $c_i$ and $B_i$ in the spectral polynomial are uniquely defined \cite{constants}. In order to see explicitly how this scheme works, various solutions of the Lorentz equations related the Gross-Neveu and Nambu-Jona-Lasinio models are discussed next.   \vspace{-1em}

\section{Electromagnetic solutions}\label{sec3}

\begin{figure}[h!]
	\centering
	\includegraphics[scale=0.16 ]{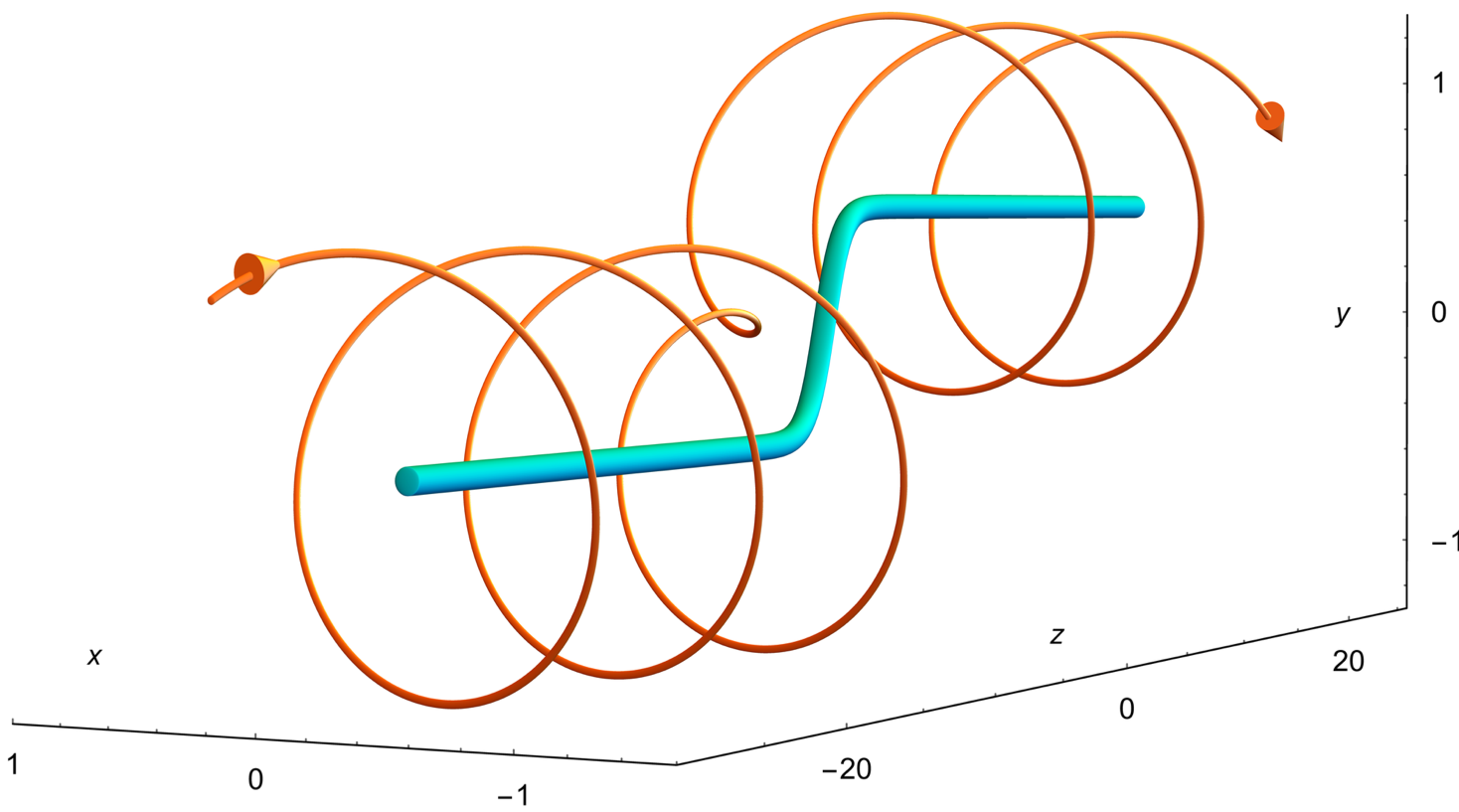}
	\caption{Trajectory (orange) for a particle under electric kink (\ref{kink}) enhanced by a cyclotronic motion (\ref{enh}) for $m=-0.75, B=4, \omega=0.65$. The kink (cyan) is shown to illustrate how the particle changes its asymptotic motion.}
	\label{fig1}
\end{figure}

{nS}$^0:$\emph{Chiral equation.} The first member in the hierarchy of NLS equations has been deeply studied in the context of field theories and gravity \cite{solitons, sdgravity}. Its stationary solutions are rather simple, but the electromagnetic counterparts and their significance are not trivial at all. The electric fields solutions of (\ref{chiral})  can be written as,
\begin{equation}\label{elec1}
(E_1,E_2)=\frac{\epsilon}{\omega}\left(\cos \frac{\omega\, m }{B-m} x^0,\,\sin \frac{\omega\, m }{B-m} x^0 \right)\, ,
\end{equation} 
where the explicit form of the coordinates is determined integrating the resolvent components (\ref{resolvent}), leading to  $(x^1,x^2)=(E_2/m,-E_1/m)$. The constants $\epsilon$, $m$ and $\omega=\sqrt{(B{-}m )^2{-}\epsilon^2}$ define the normalization condition $\eta^\mu\eta_\mu=1$. The particle describe a cyclotron motion (or an helix considering the $x^3$ case) of constant radius where both curvature and torsion are constants  ${\cal K}=m \epsilon/\omega$, ${\cal T}=m (B-m )/\omega$. 
Naturally, it is not surprising to find such solutions in the literature in (non)-relativistic scenarios \cite{books2, plasma}, usually for constant fields instead of time-independent configurations. A deeper interpretation of the solutions (\ref{elec1}) arises from the fact that the whole hierarchy is covariant under the transformations
\begin{equation}\label{enh}
\widetilde{\mathcal{E}}(\tau)=e^{i \Omega \tau}\mathcal{E}(\tau)\, ,
\end{equation}
which means that any trajectory can be enhanced with a cyclotron motion of frequency $\Omega$ as a new independent solution of (\ref{lorentzp}) inducing a completely different behaviour, see  the next example and FIG \ref{fig1}.

\begin{figure}[h!]
	\centering
	\includegraphics[scale=0.335]{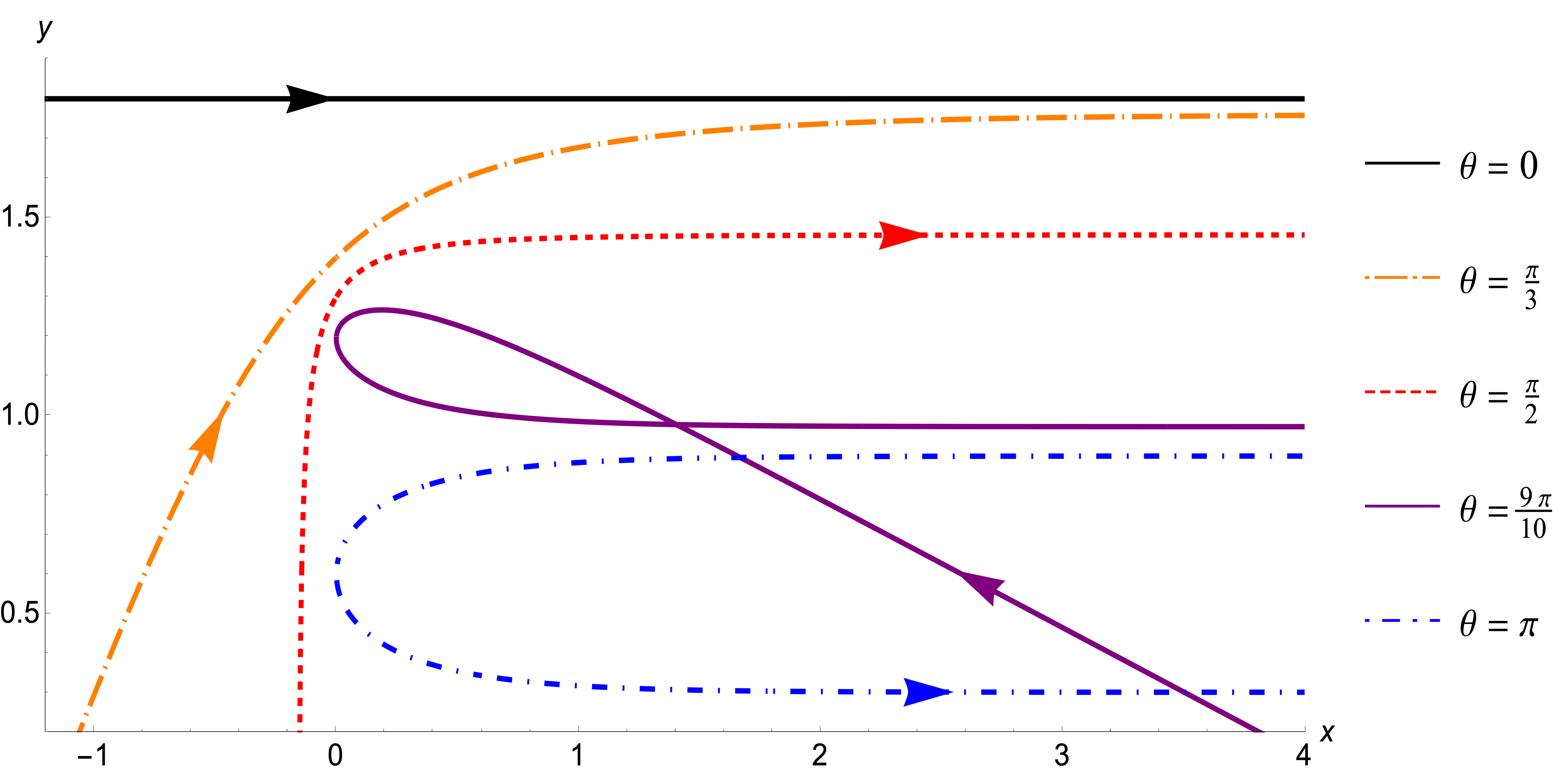}
	\caption{Motion in the plane of a particle under the electric field of the form of the Shei kink (\ref{shei}) for different values of the angle $\theta$ for $B=3$ and $m=1$. The case $\theta=\pi$ corresponds to the CCGZ kink (\ref{kink}). The trajectories have been shifted vertically for comparison purposes.}
	\label{fig2}
\end{figure}

\begin{figure}[h!]
	\centering
\includegraphics[scale=0.18]{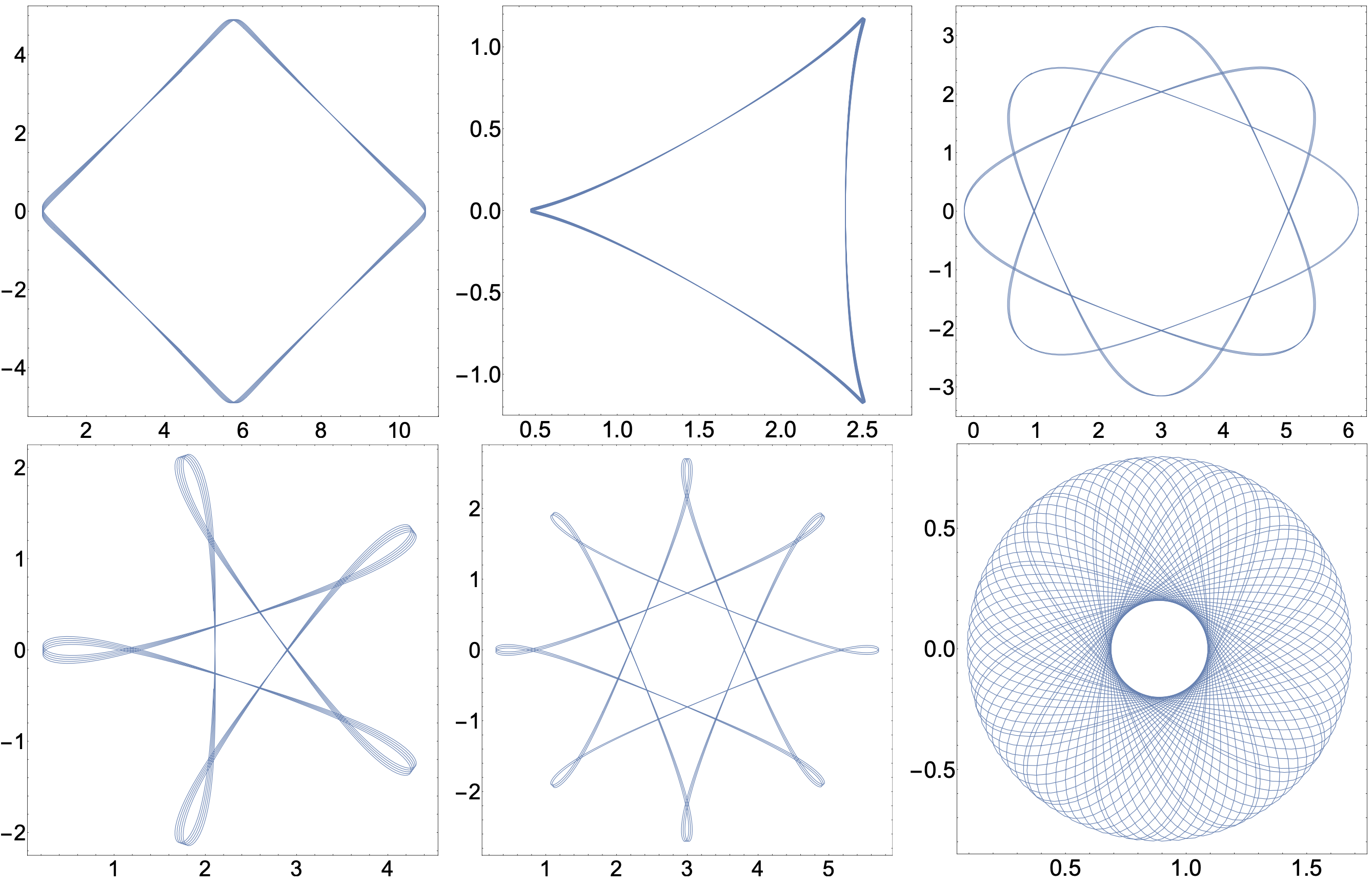}
\caption{Trajectories in the $x-y$ plane for $B=2.5$ and different values of the modulus parameter $\nu$ and angles $\theta=r\, {\bf K}^\prime(\nu)$ using the notation $(\nu,r)$. From left to right the pair or parameters are $(0.9,1), (0.45,1.3), (0.9,2.5)$ (upper panel) and  $(0.85, 1.6), (0.9, 1.5), (0.3, 1.8)$ (lower panel). The trajectories are plotted at time intervals to show their pattern, which are rotated after a few cycles. }
	\label{fig3}
\end{figure}

{nS}$^1:$\emph{NLS equation}. This equation contains several types of solutions that induce different types of motion for a charged particle. In the integrable scheme of the Lorentz equation, the curvature and torsion can be written in general form using the integrals $g_\ell$ in (\ref{recg}),
\begin{align} \label{curvature}
{\cal K}&=\sqrt{\alpha^2\left(\delta+2\beta+|{\cal E}|^2 \right)^2-4\beta} \, , \\
\label{torsion}
{\cal T}&=\frac{k_1}{{\cal K}^2}-\left(1-\frac{k_2}{{\cal K}^2} \right)\sqrt{{\cal K}^2+4\beta} \, .
\end{align}
The parameters  $\alpha=2{\cal N}_1(B)(B{+}2c_1), \beta=\frac{4 \left(c_2^2+2 c_4\right)}{\left(B+2 c_1\right){}^2}, \delta=\frac{2 \left(B c_2+2 c_3\right)}{B+2 c_1}, \rho=B c_1+2 c_2, k_1=4 \alpha  \beta  (\rho{-}4 \beta{-}3 \delta)$ and $k_2=2 \left(4 \beta -2 c_2+\delta \right)$ are given in terms of the constants $\{c_1,c_2,c_3,c_4\}$ \cite{constants}. One of the simplest solutions of the stationary NLS equation (\ref{nls}), known in the Gross-Neveu models as the Coleman-Callan-Gross-Zee (CCGZ) kink solution \cite{integrableGN, bdg1, sols}, in terms of the constant $m$ and the electric fields reads
\begin{equation}\label{kink}
E_1(\tau) = m \tanh m\tau  , \quad E_2(\tau) = 0 \, .
\end{equation}
The trajectories can be easily computed from the resolvent, verifying  explicitly that $x^0(\tau)$ is a single valued function of the proper time,
\begin{align}\notag%\label{}
x^\mu(\tau)=\frac{1}{B \sqrt{B^2-4m^2}} \left(
\begin{array}{c}
2 B^2 \tau -4 m \tanh m \tau \\[3 pt]
4 B \log \left(\cosh m \tau \right) \\[3 pt]
-4 m \tanh \epsilon \tau
\end{array}
\right) \, ,
\end{align}
where the integration constants were set to zero. The kink type of electric field (\ref{kink}) provokes that the particle travels from the remote past at to the remote future at $x^2_\mp=\pm\frac{ 4m}{B \sqrt{B^2-4m^2}}$. Near the origin, the particle smoothly changes its incoming and outgoing directions in $x^1(\tau)$, see FIG \ref{fig2}, according that both torsion and curvature are asymptotically constant 
\begin{equation}
{\cal K}=-{\cal T}=2m^2 \text{sech}^2 (m \tau)/ \sqrt{B^2{-}4m^2} \, .
\end{equation}
This situation resembles the standard reduction of a $\lambda \, \phi^4$ theory, with spontaneous symmetry breaking potential in $1{+}1$ dimensions, which finds stationary and inhomogeneous solutions to the equations of motion. By neglecting the time derivatives, swapping the role of the space and time coordinates, and reversing the sign of the  $\lambda$ in the potential, one ends up with a one dimensional particle moving in a potential. In this scenario, the standard kink solution, corresponds to the movement of the particle from the top of the unstable equilibrium point of the potential, in the remote past, to the other unstable point in the remote future. Using the cyclotronic covariance (\ref{enh}) and the electric fields (\ref{kink}), a new solution can be constructed in which the particle mixes both motions, i.e., moves in a helix centered on a fixed point and then changes its centre due to the kink nature of the solution, as it is depicted in FIG \ref{fig1}. The complex generalisation of the CCGZ kink is known as the Shei complex kink \cite{shei, bdg1}, a known solution of the Nambu-Jona-Lasinio model. In terms of time-dependent electric fields, the twisted kink solution is
\begin{equation}\label{shei}
{\cal E}(\tau)=\frac{m \cosh \left( m  \sin \left(\frac{\theta }{2}\right) \tau -\frac{i \theta }{2}\right) e^{\frac{i \theta }{2}} }{ \cosh \left( m \sin \left(\frac{\theta }{2}\right) \tau \right)} \, .
\end{equation}
that depends on a parameter $\theta$ interpolating two asymptotic regions ${\cal E}(\tau=+\infty)=e^{-i \theta}{\cal E}(\tau=-\infty)$, recovering the real electric field (\ref{kink}) when $\theta=\pi$. The word-lines in this case may have a loop in dependence of the angle $\theta$, see FIG. \ref{fig2}. The coordinates can be also integrated but we omit the explicit formulas here. The example of the twisted kink can be better understood with the analogy with a complex scalar field with a symmetry breaking. In this case, the particle starts out from an unstable maximum in the remote past with a non zero angular momentum and ends in other unstable maximum in the remote future but with a distinct the final direction, corresponding to the change in the complex phase of the minimum of the field. All of the previous examples admit a periodic generalisation known as complex crystal kink \cite{bdg1}, 
\begin{equation}
 {\cal E}(\tau)=\mu \frac{\sigma(\mu\, x+i{\bf K}^\prime
    -i\theta/2)}{\sigma(\mu\, x+i{\bf K}^\prime)\sigma(i\theta/2)} e^{i \lambda x+i\,\theta \zeta(i{\bf K}^\prime) /2} \, ,
\label{periodic}
\end{equation}
in terms of the periodic Weierstrass zeta and
sigma functions and the parameters $\mu=-2im\, {\rm sc}\left(i\theta/4\right) {\rm
nd}\left(i\theta/4\right)$ and $\lambda=\mu\left(-i\,\zeta(i\theta/2)+i\,{\rm
    ns}(i\theta/2)\right)$ \cite{ellip, expla}. As the curvature and torsion are given in terms of electric field modulus $|{\cal E}|$ (\ref{curvature}) and (\ref{torsion}), the motion follow a non-trivial cyclotron type of motion according to the period $P=2 {\bf K}^\prime(\nu)/\mu$. This period produces a constant phase on the electric fields $ {\cal E}(\tau+P)=e^{2i\varphi}{\cal E}(\tau)$ and the trajectories are not closed in the generic case, see FIG. \ref{fig3}.

\begin{figure}[h!]
	\centering
\includegraphics[scale=0.16]{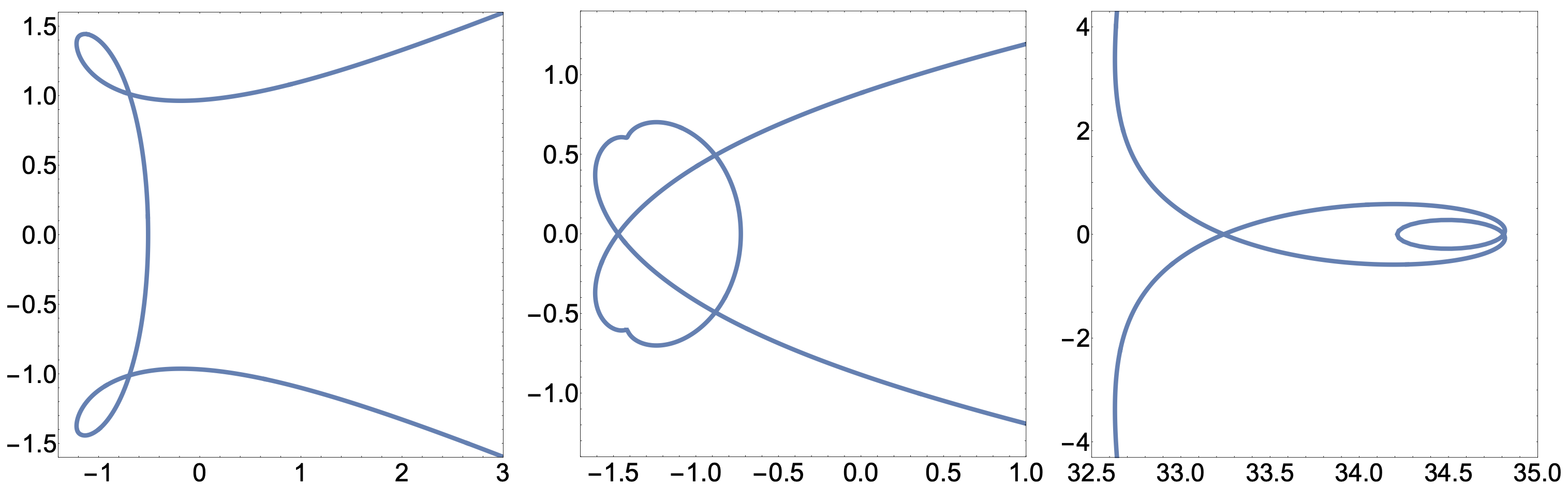}
\caption{Trajectories in the $x-y$ plane for the twisted two kink (\ref{kink}), showing the different two loops, for $B=2.5$, $\gamma_1=i$, $\gamma_2=0$, $\theta_2=\pi/6$ and $\theta_1=1.76 \pi$ (left), $\theta_1=1.6 \pi$ (center) and $\theta_1=1.18 \pi$ (right). }
	\label{fig4}
\end{figure}

{nS}$^2:$ \emph{Hirota and mKdV equation}.  The last example is the two-parameter generalization of the Shei complex kink (\ref{shei}), defined via $\tau_\ell=m \tau \sin \theta_\ell +\gamma_\ell$ with the form
\begin{equation}\label{twokink}
{\cal E}\!=\! i m \, \sech \tau_1 \,\, \sech \tau_2\,\frac{\cos\Theta_+\sin \theta_-{-}\cos\Theta_-\sin \theta_+}{2 \frac{\cos \theta_1/2}{\cos \theta_2/2}-\frac{\sin \theta_1/2}{\sin \theta_2/2}\, \tanh \tau_1 \, \tanh \tau_2} \, ,
\end{equation}
where constants are $\Theta_\pm=2\theta_{\pm}-i(\tau_1{\pm}\tau_2)$ and $\theta_{\pm}=\frac{\theta_1\pm\theta_2}{2}$ \cite{cj}. The twisted electric field in the remote past and future behaves as ${\cal E}(\tau=+\infty)=e^{-i (\theta_1+\theta_2)}{\cal E}(\tau=-\infty)$ and the trajectories have an incident and scattering angle depending on the two angles $\theta_1$ and $\theta_2$, where each of them contributes with a loop, see FIG. \ref{fig4}. A special case of (\ref{twokink}) is the kink-antikink Dashen-Hassler-Neveu baryon \cite{integrableGN, sols}  $\theta_1=\theta_2-\pi$, where the electric field component $E_1$ vanishes and takes the form $E_2(\tau)=\coth b
-\tanh \left(\tau+b/2\right)+\tanh \left(\tau-b/2\right)$, which satisfies the mKdV equation. In this case the world line simply follows the same incident direction after making a complete loop. 
 \vspace{-1.5em}
\section{Conclusion}\label{sec4}
In this article we have established a correspondence between the condensate configurations of two-dimensional fermion systems and the trajectories of a relativistic charged particle in three and four dimensions via the Dik'ii-Eilenberger equation. Thanks to the underlying integrable nonlinear Schr\"odinger hierarchy, a direct method has been developed to construct an infinite number of solutions of different nature. Since most of these solutions have not yet been explored, some next natural candidates are kink solutions moving in a non-trival (elliptic) background, $n$th twisted complex kinks, rational and degenerate solitons, together with their combinations. Furthermore, the scenario of possibilities is huge, since the solutions can always be enhanced with (\ref{enh}) to cyclotron-like motion. The examples shown here demonstrate the variety of trajectories that can be designed by suitable electric fields, which can be useful for certain special conditions, such as the longitudinal electric field that plays an interesting role in plasma physics \cite {cj}. The time dependence of the electric field directly establishes that the usual Lorentz electromagnetic invariants are not preserved in this case. However, since the electric fields satisfy the nonlinear integrable equations $\text{nS}^m$, the recursive relations (\ref{recg}) provide an infinite number of constants $c_\ell$ with $\ell>m$ in terms of the electromagnetic field and trajectories. These conserved quantities have no independent geometric interpretation, but both torsion and curvature are given in terms of them. It is still interesting to study whether these quantities may play a role from other integrable approaches \cite{hiint}.
 \vspace{-1.65 em}

\section*{Acknowledgements}
\vspace{-1em}
FC was supported by Fondecyt grants 1171475 and 1211356. FC thanks the kind hospitality of Universitat de Barcelona. JLS thanks the Spanish Ministry of Universities and the European Union Next Generation EU/PRTR for the funds through the Maria Zambrano grant to attract international talent 2021 program.


\begin{thebibliography}{99}
\bibitem{porsi}{Given a trajectory one always can recover the electric field in terms of the the original time by inverting the relation $x^0=x^0(\tau)$. This is always possible by virtue of the normalization condition for the covariant velocity $\eta^\mu$ because, $\eta^0=\frac{dx^0}{d\tau}>0$ from the inverse relation $\tau=\tau(x^0)$.}
\bibitem{constants} The constants $c_\ell $  are fixed in terms of the parameters $B_i$ which are the band edge energies or bound states of the BdG Hamiltonian via the relation
$c_k=\sum_{\substack{i=j_0,j_1,...,j_n=0 \\ j_0+j_1+..+j_n=k}}^{k} 2^{-2k}
\prod_{i=0}^{2n+1} \frac{(2 j_i )!}{(j_i !)(2 j_i-1)} B_i^{j_i}$
where $k=1,..., n+1$ and $c_0=1$.
\end{thebibliography}
\end{document}